\documentclass{kluwer}
\usepackage{epsfig}
\begin{document}
\begin{article}
\begin{opening}

\title{Counts, Sizes and Colors of \\ Faint Infrared Selected Galaxies}

\author{Paolo \surname{Saracco}\email{saracco@merate.mi.astro.it}}
\institute{Osservatorio Astronomico di Brera, Merate, Italy}
\author{Sandro \surname{D'Odorico}\email{sdodoric@eso.org}}
\author{Alan \surname{Moorwood}\email{amoor@eso.org}}
\author{Jean G.\surname{Cuby}\email{jcuby@eso.org}}
\institute{European Southern Observatory, ESO, Garching, Germany}

\runningtitle{IR Faint Galaxies}
\runningauthor{P. Saracco}

\begin{ao}
Paolo Saracco\\
Osservatorio Astronomico di Brera\\
via E. Bianchi 46, 22055 Merate, Italy\\
e-mail: saracco@merate.mi.astro.it
\end{ao}

\begin{abstract}
Deep J and Ks band images covering a $5\times5$ arcmin
area    centered on the NTT Deep Field  have been obtained  during the
Science Verification of SOFI at the NTT. 
These images were made available via the Web in early June. 
The preliminary results we have obtained by  the analysis of these data
are the following:
{\em i)} the counts continue to rise  with no evidence of a turnover 
or of a flattening down to the limits of the survey (Ks=22.5 and J=24); 
{\em ii)} we find a slope $dlog(N)/dm\sim$0.37 
in agreement with most of the faintest surveys but much steeper than 
the Hawaii survey;
{\em iii)} fainter than Ks$\sim19$ and J$\sim20$, the median J-K color 
of galaxies shows a break in its reddening trend turning toward bluer colors;
{\em iv)}  faint bluer galaxies also display a larger compactness index, 
and a smaller apparent size.

\end{abstract}
\end{opening}

\section{Observations and Data Reduction}
Observations in J and Ks bands have been carried out at the 
ESO 3.5 m New Technology Telescope (NTT) with the new infrared 
imager/spectrometer SOFI (Moorwood et al. 1998).  
It is equipped with a 1024$\times$1024 pixel Rockwell Hawaii array
(0.292 arcsec/pix).
The observed field, centered at $\alpha=12^h:05^m:26.02^s$, 
$\delta=-07^o:43':26.43''$ (J2000) include the NTT Deep Field
(Arnouts et al. 1999).
The observations have been performed 
with  seeing conditions ranging from 0.6-0.9 arcsec 
(median FWHM = 0.75 arcsec).
A total exposure time of 624 min in Ks and 256 min in J 
was  achieved as a result of the jittered frame co-addition.

Raw data have been  corrected for dark-current pattern 
by subtracting for each night a median dark and for pixel-to-pixel 
gain variations using a differential dome flat field.
After flat fielding and standard sky subtraction, each basic set of  frames 
has been registered and finally co-added to produce the final image 
The whole procedure has been accomplished relying on the
original {\it Jitter} package by Devillard (1998; http://www.eso.org/eclipse).
\begin{table*}
\caption{ Summary of observations and image quality.}
\begin{center}

 \begin{tabular}{cccccc}

\hline
\hline
        Filter &  Number of & t$_{exp}$& $<FWHM>$& $\mu$&m$_{lim}$\\
        & frames    &  (s)   &(arcsec)&mag/arcsec$^2$& (3$\sigma$)\\
\hline
          Ks &  624 & 37440 & 0.75& 23.95&22.76\\
          J  &  128 & 15240 & 0.75& 25.75&24.57\\
\hline
\hline
\end{tabular} 
\end{center}
\end{table*}
The limiting surface brightness as well as the 3$\sigma$ magnitude within  a
2$\times$ FWHM (1.5 arcsec) circular aperture estimated on the final images
are reported in Tab. 1.

\section{Object Detection and Magnitudes}
The extraction and the magnitude estimate of sources have been 
performed with the SExtractor image analysis package (Bertin \& Arnouts, 1996).

The object detection has been performed in the central deepest portion 
(4.45$\times$4.45 arcmin) of the images by first convolving them with a 
gaussian function having a FWHM=2.5 pixels ($\sim0.73$ arcsec).
A 1$\sigma_{bkg}$  detection threshold and a  minimum detection area 
equal to the seeing disk have been used.

This rather low detection threshold provides a raw catalog from which 
we have extracted  different samples by varying the S/N 
limit in order to optimize completeness  and number of spurious detections
 for a given magnitude limit.
For example, at a $S/N=5$ detection limit, compact objects as faint 
as $Ks = 23.0$ and  $J = 24.8$ were singled out in our images, 
compared  with $Ks = 23.9-24.5$ and $J=24.9-25.2$  in the deepest 
Keck observations (Bershady et al. 1998; Djorgovski et al. 1995).
We are therefore confident that our sample is not biased against 
the detection of small compact sources, being fully comparable 
with the deepest data previously 
obtained by other authors on smaller areas on the sky
(Gardner et al. 1993; Cowie et al. 1994; Soifer et al. 1994;  
McLeod et al. 1995; Djorgovski et al. 1995; Moustakas et al. 1997; 
Minezaki et al. 1998).

We have measured the reliability of the detections by evaluating
the number of spurious sources and their magnitude distribution using
a small set of  noise images obtained
by combining  the corresponding set of frames with ``wrong'' shifts.
The  images thus obtained do not contain sources and are 
characterized by the same mean background noise of the original data.
We have deduced that  a   S/N$\ge$5 cutoff seems to ensure the 
optimum threshold for confident object detection both in J and Ks
 within a 0.3 mag accuracy.
On the basis of the above selection criterium our subsequent 
analysis will rely on the {\it bona fide} Ks and J samples 
containing 1025 sources down to Ks$\le22.5$ and 
1569 sources down to J$\le$24.0, respectively.

Magnitudes have been estimated within a diameter aperture of 2.5 arcsec
($\sim3\times$FWHM) and corrected to ``total'' using an aperture correction
of 0.25 mag obtained using a sample of bright objects.

\section{Results}
\subsection{Differential Galaxy Counts}
In order to derive  differential galaxy counts, we have first ``cleaned'' 
the two samples of stars. 
We have defined as
stars those sources  brighter than Ks$=19$ (in the Ks-band selected sample)
and  J$=20$ (in the J-band selected sample), having a 
value of SExtractor ``stellarity'' index larger than 0.9 both in the J sample
and in the Ks one.
\begin{figure*}
\centerline{\psfig{figure=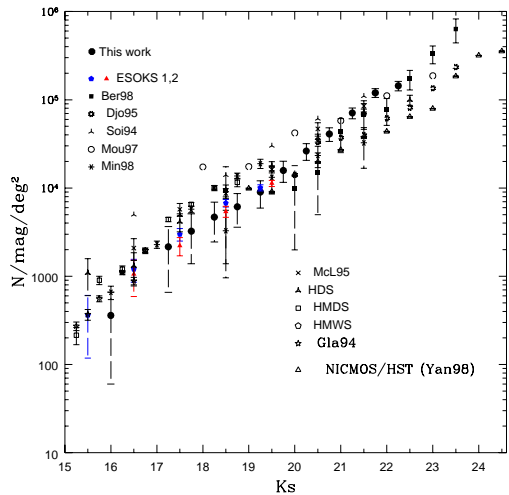,height=70mm}}
 \caption{The figure compares the Ks-band counts obtained in this 
work with those in the literature: 
  Djorgovski et al. (1995, Djo95), Bershady et al. (1998, Ber98) 
(obtained at the Keck telescope), Gardner et al. (1993, HMWS, HMDS, HDS), 
Glazebrook et al. (1994, Gla94), Soifer et al. (1994, Soi94), 
McLeod et al. (1995, McL95), Moustakas et al. (1997, Mou97),
Saracco et al. (1997, ESOKS1, 2) and Minezaki et al. (1998, Min98) 
NICMOS/HST H-band counts of Yan et al. (1998, Yan98) are also shown.}
\end{figure*}

Then we have evaluated the completeness 
correction to apply to the raw data to account for the number of 
sources undetected at a given magnitude.
This correction is obviously stronger at faint magnitudes and 
it mainly depends on the source spatial structure.
Thus, we have generated a set of frames by dimming the final J and Ks frames 
themselves by various factors while keeping constant the background noise.
This way we have used the whole real sample of sources reproducing
the whole range of size and shape.
SExtractor was then run with the same detection parameters to search for
sources in each dimmed frame.

Fig. 1 shows the Ks-band galaxy counts here derived superimposed to those in 
the literature.
Our counts follow a d$logN/dK$ relation with a slope of 0.38 in the
magnitude range $17<K<22.5$ and do not show any evidence of a turnover or
of a flattening down to the limits of the survey.

In Fig. 2 our J-band galaxy counts are compared with those of Bershady et al.
(1998), the only J-band data available in the literature.
We obtain a slope of J-band counts $\sim$0.36 in the magnitude
interval $18<J<24$.
Also the J-band counts do not show any hint of a decrease down to the
limits of the data.   
\begin{figure}
\centerline{\psfig{figure=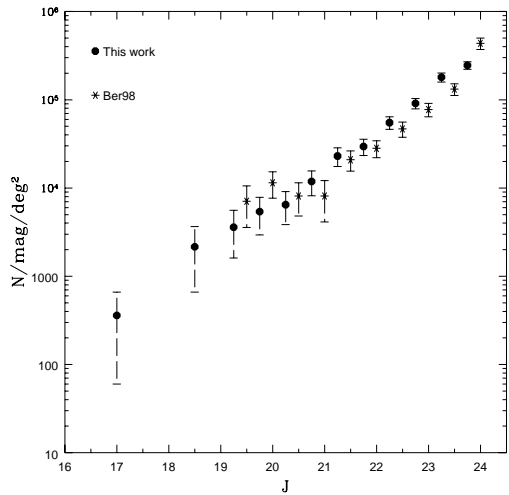,height=70mm}}
 \caption{The J-band galaxy counts obtained in this work are
compared with those previously obtained by Bershady et al. (1998, Ber98):
the counts continue to rise with a power law slope of  $\sim$0.36.}
\end{figure}

The total numbers of galaxies represented in Fig.1 and 2 are 874 for 
the Ks-band and 1285 for the J-band and represent the largest samples 
at these depths.

\subsection{Color and Size of Galaxies}
The 2.5 arcsec aperture J-K color of galaxies in our sample was obtained
by running SExtractor in the so-called {\em  double-image mode}:
J magnitudes for the Ks  selected sample have been derived
using the Ks frame as  reference image for detection and the J image 
for measurements only, {\it vice versa} for the J selected sample.

In Fig. 3  and 4  the J-Ks color of each galaxy 
is shown as a function of the apparent magnitude for the two samples.
The median J-Ks color in each magnitude bin is also over-plotted as 
filled circles.
The error bars are the standard deviation from the mean of the values
within each bin.
\begin{figure}
\centerline{\psfig{figure=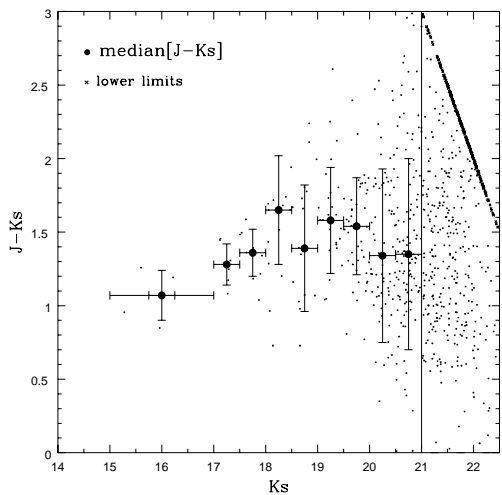,height=70mm}}
 \caption{J-Ks color of K-band selected galaxies  as a function of 
Ks magnitude. The filled symbols represent the median J-Ks color
of galaxies in each magnitude bin.
Vertical error bars are  the standard deviation from the mean of the values
within each bin.
The width of the bins are represented by the 
horizontal error bars
The small crossed symbols, distributed along the right hand line,
 represent the sources undetected in the J frame, i.e. the J-Ks lower  
limits.
}
\end{figure}
Both the samples show first a reddening 
trend of the median J-Ks color down to Ks$\sim19$ and J$\sim20$ respectively.
Fainter than these magnitudes the J-K color of galaxies follows
an unexpected behavior describing  a bluing trend  with increasing
apparent magnitudes.
\begin{figure}
\centerline{\psfig{figure=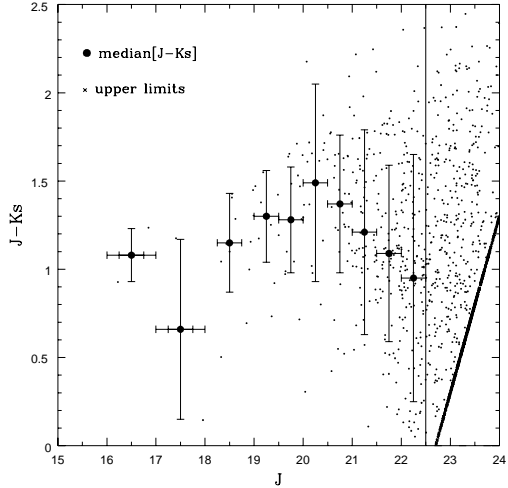,height=70mm}}
 \caption{J-Ks color of J-band selected galaxies  as a function of 
J magnitude. Symbols are as in Fig. 3.
}
\end{figure}

In order to measure the size of galaxies of our samples, we have made
use of the metric size function $\eta(\theta)$ introduced first 
by Petrosian (1976)
\begin{equation}
\eta(\theta)={{1~d~ln~l(\theta)}\over{2~d~ln(\theta)}}
\end{equation}

(Kron 1993), where $l(\theta)$ is the growth curve.
This function has the property 
$\eta(\theta)={{I(\theta)}/{\langle I\rangle}_\theta}$
where  $I(\theta)$ is the surface brightness at radius
$\theta$ and ${\langle I\rangle}_\theta$ is the mean surface brightness 
within $\theta$.

The function $\eta(\theta)$ has been obtained for each galaxy
by constructing its own intensity profile.
Following Bershady et al. (1998) we have defined the angular size 
 of galaxies the value $\theta_{\eta}$ such that $\eta(\theta_{\eta})=0.5$.
The median  $\theta_{0.5}$ values  are 0.73 arcsec and 0.79 arcsec
for the J-band and Ks-band selected sample respectively.
We have then measured the compactness of galaxies by defining the
luminosity concentration index $C_\eta$
\begin{equation}
C_\eta={{F(<\theta_{0.5})}\over{F(<1.5\theta_{0.5})}}
\end{equation}
that is the ratio between the flux within the radius  $\theta_{0.5}$
and the flux interior to the radius $1.5\theta_{0.5}$.
In Fig. 5 the concentration index $C_\eta$ and the apparent radius
$\theta_{0.5}$ of galaxies belonging to the Ks sample are shown
as a function of the apparent magnitude Ks in the upper and lower panel
respectively.
\begin{figure}
\centerline{\psfig{figure=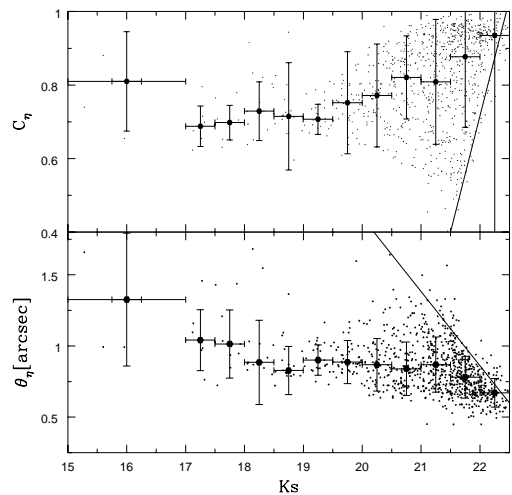,height=70mm}}
 \caption{Light concentration index $C_\eta$ (upper panel) 
and apparent size $\theta_{0.5}$ (lower panel) of galaxies as a function 
of their Ks magnitude.
The filled circles are the median values in 0.5  magnitude width bin.
Vertical error bars are  the standard deviation from the mean of the values
within each bin while horizontal  bars represent the width 
of the bins.
The solid line in the lower panel shows the magnitude enclosed in 
an aperture of radius $\theta$ and a uniform surface brightness 
Ks=22.76 mag/arcsec$^2$.
}
\end{figure}
The light concentration index seems to remain constant down to
Ks$\sim19$ and J$\sim20$ while it is evident that 
galaxies become systematically more compact to fainter magnitudes.
This trend is clearly present in both the samples as shown by the 
decrease of the median  values $C_\eta$ with increasing apparent magnitudes.
It is worth noting that the increasing compactness of galaxies
occurs in the same magnitude ranges where galaxies get bluer.
Moreover this trend seems to be matched also by a decrease of the 
apparent size of galaxies.

\section{Conclusions}
 The main results obtained by this preliminary analysis of the data 
are the following:
\begin{itemize}
\item number counts follow a  $d~log(N)/dm$ relation with a slope 
of 0.38 in Ks and of 0.36 in J showing no sign of flattening or 
turnover down to the faintest magnitudes. 
Such slopes and behavior fully agree with most of the ground-based 
data and with the deepest NICMOS/HST data (Yan et al. 1998); 

\item fainter than Ks$\sim19$ and J$\sim20$, the median J-K color 
of galaxies shows a break in its reddening trend turning toward bluer colors;

\item faint bluer galaxies also display a larger compactness index, 
and a smaller apparent size.

\end{itemize}
The absence of a turnover in the counts down to the faintest magnitudes
is indicative of an increasing contribution of sub-L$^*$ and hence
local ($z \ll 1$) galaxies leading to  a substantial 
steepening in the  faint-end tail of the galaxy IR LF. 
Such a claim is also supported by the observed color trend which 
seems to rule out any major contribution of high-$z$ galaxies as 
primary contributors to faint counts. 

The existence of a steep  ($\alpha\ll-1$) faint-end tail ($L<0.1-0.01L^*$)
in the IR LF would naturally account both for the J-K color trend and
the compactness trend shown by
 galaxies.

\end{article}

\end{document}